\documentclass[12pt]{iopart}
\usepackage{iopams}

\usepackage{epsfig}
\usepackage[usenames]{color}
\usepackage{soul}
\usepackage{graphicx}
\usepackage[hypertex]{hyperref}
\usepackage{srctex}

\usepackage{dcolumn}
\usepackage{bm}

\begin{document}

\title{Internal dipolar field and soft magnons in periodic nanocomposite magnets}

\author{A.~M.~Belemuk$^{1}$, S.~T.~Chui$^{2}$}
\address{$^1$ Institute for High Pressure Physics, Russian Academy of Science, Troitsk 142190, Russia}
\address{$^2$ Department of Physics and Astronomy and Barto
Research Institute, University of Delaware, Newark, Delaware 19716, USA}

\begin{abstract}
We study spin wave excitations in a three-dimensional
nanocomposite magnet of exchange coupled hard (SmCo$_5$) and soft
(FeCo) phases. The dipolar interaction splits the spin wave
energies into the upper and lower branches of the spin wave
manifold. When the amount of the soft phase is increased the
energy of low-lying spin excitations is considerably softened due
to two reasons: (i) the low- lying mode locked into the soft phase
region with a spin wave gap at ${\bf k}= 0$ which scales
approximately proportional to the anisotropy constant of the soft
phase and (ii) the internal dipolar field which comes from
magnetic charges forming at hard-soft boundaries with normals
parallel to the magnetization displaces the spin wave manifold
toward the lower energies. With adding more soft phase the spin
wave gap closes and the system moves to another ground state
characterized by the magnetization mismatch between spins of the
hard and soft phases.

\end{abstract}


%
\vspace{2pc}

%

\section{Introduction}

Materials with periodically modulated magnetic and geometric
properties are of special interest recently from the viewpoint of
applications, which aim to manipulate propagating spin waves
\cite{Serga10, Chumak09, Lenk11, Pirro14, Jungfleisch15}. Spin waves propagating in
nonhomogeneous magnetic nanostructures serve as information
carriers and show the existence of allowed frequency ranges and
forbidden band gaps \cite{Gubbiotti05, Wang09, Bauer97, Zivieri12,
Gubbiotti12, Tacchi11}. The periodic modulation of magnetic
properties are also realized in nanocomposite magnets composed of
exchange-coupled hard and soft magnetic phases \cite{Victora05,
Kneller91, Fullerton98}. The hard phase provides the immense
magnetic anisotropy that stabilizes the exchange-coupled soft phase
against demagnetization. In multilayer geometry it gives the
increase in the remanent magnetization and the ultimate gain in
the energy product with increasing amount of the soft phase
material \cite{Skomski93}. However, in a geometry where the
nanocomposite is magnetized perpendicular to the hard-soft
boundary, the anticipated increase in the remanence does not occur
partly because the hard and soft phase magnetization vectors are
never completely parallel to each other \cite{Chui12}.

In a nonhomogeneous magnetic material values of the saturation
magnetization $M_S$, anisotropy constant $K$ and the direction of
the magnetization ${\bf n}_0$ are, in general, functions of the
position vector ${\bf r}$. In a magnonic crystal the applied
uniform magnetic field usually forces all the magnetic moments to
be magnetized in the direction of the applied field
\cite{Krawczyk08, Vohl89, Tiwari10}. In a nanocomposite hard-soft
magnet the behavior of the magnetization vector ${\bf M}({\bf r})=
M_S({\bf r}) {\bf n}_0({\bf r})$ depends on the demagnetizing effects and
mutual arrangement of easy axes of constituent ferromagnets. For
an arrangement with easy axis (in the $z$ direction)
perpendicular to the hard-soft boundary the homogeneously
magnetized state ${\bf M}({\bf r})= M_S({\bf r}) {\hat z}$ is
energetically unfavorable from the magnetostatic point of view.
Because of the discontinuity of the magnetization at the hard-soft
boundary, a magnetic charge $\rho_M= -\nabla \cdot {\bf M}$ is
developed. This increases the dipolar energy which can be
written in the form~\cite{Aharoni2000} $E_{dip} \propto \int d^3r d^3r'\rho_M({\bf
r})\rho_M({\bf r'})/|{\bf r}- {\bf r'}|$. For sufficiently
low soft phase content a strong anisotropy field of the hard phase
and exchange forces at soft-hard boundaries enforce the whole
magnet to be magnetized in the $z$ direction. With adding more
soft phase with considerably smaller value of anisotropy constant
the soft phase spins become tilted from the easy direction and
their averaged direction are misaligned with spins of the hard
phase. Monte Carlo simulation at finite temperatures
\cite{Chui13} reveals that this misorientation grows with
temperature and as the amount of the soft phase is increased. Such
a misalignment, usually, is not considered in the context of magnonic
crystals where the nonuniform static demagnetizing filed is
assumed to be homogeneously averaged throughout the sample
\cite{Krawczyk08}.

The dependence of the ground state  magnetization in composite
permanent magnets on the demagnetizing effects has been discussed
previously for a single soft inclusion in a matrix of hard phase
\cite{Kronmuller98}. These calculations reveal that the remanent
magnetization sensitively depends on the size of the inclusion.
With increasing of the fraction of the soft phase the long-range
stray field destroys the parallel alignment of the soft magnetic
moments and creates magnetic vortex-like structures. These
calculations are based on computational micromagnetism
\cite{Kronmuller96, Schrefl94, Fisher95} and yield a stable
magnetization distribution by minimizing the total energy of
magnetic system.

Micromagnetic calculations \cite{Kronmuller98} as well as an
analytic estimate of the nucleation field and the remanence
enhancement \cite{Skomski93} have not discussed the thermally
activated tilting of spins and the resulting magnetization
mismatch. The effect of thermal activation can be understood if
one knows the low-lying spin wave energies $E_j({\bf k})$. The
fluctuation of the magnetization is determined by the density of
the magnons which in turn is governed by the Boltzmann factor
$\exp(-E_j({\bf k})/k_BT)$. The lower the energy of spin
excitations, which turns out to be concentrated in the soft phase,
the more fluctuations of the soft phase spins and more reduction
in the remanent magnetization of the composite appears. Due to the fluctuation
of soft phase spins the remanence does not increase proportional
to the fraction of the soft phase, $M_r= v_s M_s+ v_h M_h$ ($v_s+ v_h= 1$), as it can
be for a composite with homogeneous magnetization.

To further elucidate the physics involved in the instability of the homogeneously magnetized state,
${\bf M}({\bf r})= M_S ({\bf r}) \hat z$, in the present work we
consider the spin wave spectrum in a three-dimensional (3D)
composite composed of a periodic array of hard phase cubes
immersed into a soft phase matrix. We incorporate the effect of
the nonhomogeneous saturation magnetization $M_S({\bf r})$ and
internal magnetic chargers into the formalism of spin wave
excitations.  We explicitly construct operators of spin-wave
excitations and calculate the corresponding eigenfrequencies. We
anticipate that the dipolar interaction will lower the spin wave
energy, which in turn enhances the fluctuation of the
magnetization at finite temperatures as these spin waves are
thermally excited.

A brief analysis of the spin wave spectra in a hard-soft composite was reported earlier in Ref. \cite{ChuiJAP13}
for the case of homogeneous exchange interaction, $J_{ij}= const$ for any nearest neighbors $i$ and $j$, which does not include the discontinuity of $J_{ij}$ at hard-soft boundary. That study was focused on the comparison of the behavior of low-lying spin excitations with our previous Monte Carlo simulation results \cite{Chui12}.
In this paper we extend our previous analysis
of homogeneous exchange interaction \cite{ChuiJAP13} for the case
of position-dependent exchange interaction $J({\bf R}_i, {\bf
R}_j$) and present the clarifying details of incorporating the dipolar part of interaction into the formalism of spin-wave excitations for nonhomogeneous ferromagnets.
The main purpose of the work is to extract the effective demagnetizing field $H_{dip}$ from the spin-wave
dispersion behavior and discuss the implication of $H_{dip}$ for spin-wave manifolds in
two-phase and one-phase ferromagnets.

The paper is organized as follows. In Section II we present the general theory of linear spin-waves in a nonhomogeneous two-phase
periodic structure of exchange-coupled hard and soft phases. We first linearize the spin Hamiltonian in the position space and then perform the diagonalization in the Fourier space.
We then proceed in comparing the analytical results obtained for the behavior of low-lying spin excitations
in a nonhomogeneous composite with the spin-wave spectra for homogeneous ferromagnets.
In Sec. III we discuss results of our calculation and the effect of the internal
demagnetizing field on the spin wave dispersion of the two-phase magnet. And Sec. IV contains our conclusion.

\section{Theoretical model}

\subsection{Two-phase composite magnet}

\begin{figure}
\includegraphics[width=0.8\columnwidth]{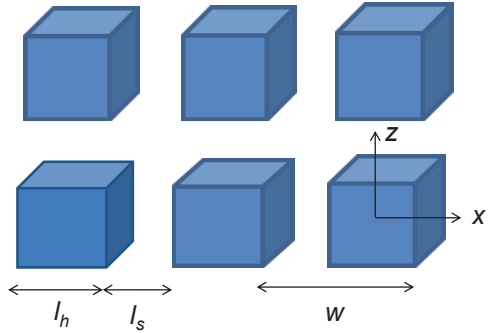}
\caption{\label{geom} Geometry of hard-soft periodic structure.
The $z$ axis is the easy axis for the hard and soft phases. }
\end{figure}

We model the hard-soft composite as a periodic array of identical
cubes of hard phase embedded into a soft phase matrix, as
illustrated in Fig. \ref{geom}. The easy axes of both phases are
in the $z$ direction. Each cube has a linear dimension $l_h$ and
separated from the adjacent one by a soft phase with linear
dimension $l_s$, so there is a periodicity in the $x$, $y$ and $z$
directions with a period $w= l_h+ l_s$. We shall refer to this
periodicity as $w$- periodicity.

We focus on the low-lying spin excitations that occur gradually
over a large distance. The magnetization, ${\bf M}({\bf r})$, for
a coarse-grained system, is defined on a discrete set of sites
${\bf R}_i$ in terms of block spin variables ${\bf S}_i=~ {\bf
S}({\bf R}_i)$, $|{\bf S}_i|=~ 1$, as ${\bf M}({\bf r}) \simeq M({\bf R}_i)
{\bf S}_i$, with $M({\bf R}_i)= M_S({\bf R}_i)$ being the saturation
magnetization density of the hard ($M_{h}$) or soft ($M_{s}$)
phase at site $i$. Spins of the hard and soft phases are arranged
in a three-dimensional cubic lattice with an effective lattice
constant $a$. The effective magnetic moment of site $i$ is $M({\bf R}_i)
v$, where $v= a^3$ is the volume of a block spin cell.

The interaction between two spins ${\bf S}_i$, ${\bf S}_j$ located
at ${\bf R}_i$, ${\bf R}_j$ is described by the Hamiltonian for a
classical spin system, including the nearest-neighbor exchange
energy, the uniaxial magnetic anisotropy term and the
dipole-dipole interaction:
\begin{equation} \label{Htot}
\fl
H= -\frac{1}{2} {\,} \mathop{{\sum}'}_{i,j} {\,} J_{ij} {\,} {\bf S}_i \cdot {\bf S}_j-
\sum \limits_i {\;} K_i {\,} \left(S^{z}_i\right)^2
- \frac{g}{2} {\,}  \mathop{{\sum}'}_{i,j} {\,}
D^{\alpha \beta}({\bf R}_{ij}) M({\bf R}_i) M({\bf R}_j){\,} S^{\alpha}_i S^{\beta}_j,
\end{equation}
where ${\bf R}_{ij}= {\bf R}_i- {\bf R}_j$, and the summations are
over all distinct magnetic sites $i$ and $j$  with the restriction
that ${\bf R}_{ij}\neq 0$. Indices $\alpha$ and $\beta$ denote the
Cartesian components $x$, $y$ or $z$, and $D^{\alpha \beta}({\bf
R}_{ij})= (3 R^{\alpha}_{ij} R^{\beta}_{ij}- R^2_{ij}
\delta_{\alpha \beta})/R^5_{ij}$ is the dipolar interaction
tensor. In Eq. (\ref{Htot}) positions vectors ${\bf R}_i$ are
given in units of the lattice spacing $a$. The exchange constant
$J_{ij}$, is equal to $J_h$ ($J_s$) for the nearest
neighbor spins of hard (soft) phase and zero otherwise. The two
phases are exchange-coupled with a coupling constant $J_{hs}$ which can be estimated as
the geometric mean of $J_h$ and $J_s$, $J_{hs}=
(J_hJ_s)^{1/2}$. $K_i$ is the anisotropy constant of the hard
($K_h$) or soft ($K_s$) phase. The characteristic magnetostatic
energies of the hard and soft phases are  $g_{h}= \mu_0M_h^2
v/(4\pi)$ and $g_s= \mu_0M_s^2 v/(4\pi)$, respectively. In Eq.
(\ref{Htot}) we assume that the magnetization density $M({\bf R}_i)$ is
normalized by the magnetization of the hard phase $M_h$, and the
dipolar coupling constant $g$ is equal to $g_h$.

We use the Holstein-Primakoff transformation \cite{Holstein40} to
express the spin operators through the boson creation and annihilation
operators, $a^{\dagger}_i$,  $a_i$, as $S^+_i \simeq
\sqrt{2S} {\,} a_i$, $S^-_i \simeq \sqrt{2S} {\,} a^{\dagger}_i$, and
$S^z_i \simeq S- a^{\dagger}_i a_i$. We assume that the ground
state is the one with all spins of both phases aligned along
the $z$ direction and expand near this aligned state.

In the harmonic approximation we can rewrite the spin Hamiltonian as
\begin{eqnarray} \label{HRij}
\fl
H= \mathop{{\sum}'}_{i,j} {\,} J_{ij} \left [ a^{\dagger}_{i} a_{i}- a^{\dagger}_{i} a_{j} \right]+
\sum \limits_i {\,} 2K_i  a^{\dagger}_{i} a_{i} + g \mathop{{\sum}'}_{i,j} {\,} M({\bf R}_i) M({\bf R}_j) \times \nonumber \\
\times \left \{  A_1({\bf R}_{ij}) a^{\dagger}_{i}
a_{j}+  A_2({\bf R}_{ij}) a^{\dagger}_{i} a_{i} + \frac12 B({\bf R}_{ij}) a_{i} a_{j} + \frac12 B^*({\bf R}_{ij}) a^{\dagger}_{i}
a^{\dagger}_{j} \right \}  ,
\end{eqnarray}
where coefficients $A_{1,2}({\bf R}_{ij})$ and $B({\bf R}_{ij})$ are
\begin{eqnarray}
&A_1({\bf R}_{ij})= \frac{1}{2}  D^{zz}({\bf R}_{ij}); \qquad
A_2({\bf R}_{ij})= D^{zz}({\bf R}_{ij}), \label{AR} \\
&B({\bf R}_{ij})=  (-1/2) \left[ D^{xx}({\bf R}_{ij})- D^{yy}({\bf R}_{ij})- 2i D^{xy}({\bf R}_{ij}) \right]. \label{BR}
\end{eqnarray}
Diagonalizing the spin-wave Hamiltonian, Eq. (\ref{HRij}), can be
performed in usual way transforming to new quasiparticle boson
operators $\alpha_l, \alpha_l^{\dagger}$, $l= 1, \dots, N$, with
$N$ being the number of lattice sites, and using the $u-v$
Bogoliubov transformation
\begin{equation} \label{uvtrR}
\eqalign{
a_i= \sum \limits_l {\,} \left\{ u_{l}({\bf R}_i) \alpha_l-
v^*_{l}({\bf R}_i)  \alpha^{\dagger}_l \right\}, \\
a_i^{\dagger}= \sum \limits_l {\,} \left\{ u^*_{l}({\bf R}_i)
\alpha_l^{\dagger}- v_{l}({\bf R}_i)  \alpha_l \right\}.}
\end{equation}
The coefficients of the transformation $u({\bf R}_i)$ and $v({\bf R}_i)$ can be found
from the equation of motion for operators $a_i$ and $\alpha_l$, provided that
$\alpha_l$ corresponds to the eigenmode of the system, $i\hbar \dot \alpha_l= E_l \alpha_l$.
From this we obtain the following system of equations
\begin{eqnarray}  \label{ABsysR1}
\fl
2K_i u({\bf R}_i)+ \sum \limits_{j \ne i} \Bigl\{ \Bigl(gA_1({\bf R}_{ij}) M({\bf R}_i) M({\bf R}_j)- J_{ij}\Bigr) u({\bf R}_j) \nonumber \\
\fl +\Bigl(gA_2({\bf R}_{ij}) M({\bf R}_i) M({\bf R}_j) + J_{ij}\Bigr) u({\bf R}_i)
- B^*({\bf R}_{ij}) M({\bf R}_i) M({\bf R}_j) v({\bf R}_j) \Bigr\}= E_l u({\bf R}_i),
\end{eqnarray}
\begin{eqnarray}  \label{ABsysR2}
\fl
-2K_i v({\bf R}_i)+ \sum \limits_{j \ne i} \Bigl\{ -\Bigl(gA_1({\bf R}_{ij}) M({\bf R}_i) M({\bf R}_j)- J_{ij}\Bigr) v({\bf R}_j) \nonumber \\
\fl -\Bigl(gA_2 ({\bf R}_{ij}) M({\bf R}_i) M({\bf R}_j)+ J_{ij}\Bigr) v({\bf R}_i)
 + B({\bf R}_{ij}) M({\bf R}_i) M({\bf R}_j) u({\bf R}_j)
\Bigr\} = E_l v({\bf R}_i).
\end{eqnarray}

To proceed further we expand the eigenmode functions $u({\bf R})$ and $v({\bf R})$
into a series on a full set of functions $\{ \varphi_l({\bf R}_i) \} $
\begin{equation}
\eqalign{
u({\bf R}_i)= \sum \limits_{l'} u_{l'} {\,} \varphi_{l'}({\bf R}_i), \\
v({\bf R}_i)= \sum \limits_{l'} v_{l'} {\,} \varphi_{l'}({\bf R}_i).}
\end{equation}
The corresponding coefficients of the expansion $u_l$ and $v_l$ can be
found from the following system of $2N$ coupled equations
\begin{equation}  \label{ABsys}
\left \{
\begin{array}{rcl}
A_{ll'} u_{l'}- B^{\dagger}_{ll'} v_{l'} &=& E_l u_l, \\
B_{ll'} u_{l'}- A_{ll'} v_{l'} &=& E_l u_l
\end{array}
\right.
\end{equation}
where summation is implied over the repeated index.
This can be conveniently rewritten in the block matrix form as
\begin{equation} \label{ABm}
\left(
\begin{array}{cc}
A & -B^{\dagger} \\
B & -A
\end{array}
\right)
\left(
\begin{array}{c}
u \\
v
\end{array}
\right)
= E
\left(
\begin{array}{c}
u \\
v
\end{array}
\right),
\end{equation}
where vector $\left (u \atop v \right)$ has components $(u_1, \dots, u_N,
v_1, \dots, v_N)^T$ and $A$ and $B$ are N-by-N matrices with the following matrix elements
\begin{eqnarray} \label{Al}
\fl
A_{ll'}= \sum \limits_i {\,} \varphi^*_l({\bf R}_i) {\,} 2K_i {\,} \varphi_{l'}({\bf R}_i)
+ \mathop{{\sum}'}_{i,j} {\,} \varphi^*_l({\bf R}_i)
\Bigl( gA_1({\bf R}_{ij}) M({\bf R}_i) M({\bf R}_j)- J_{ij} \Bigr)
\varphi_{l'}({\bf R}_j)  \nonumber \\
+ \mathop{{\sum}'}_{i,j} {\,} \varphi^*_l({\bf R}_i)
\Bigl( gA_2 ({\bf R}_{ij}) M({\bf R}_i) M({\bf R}_j) + J_{ij} \Bigr)
\varphi_{l'}({\bf R}_i),
\end{eqnarray}
\begin{equation}
\eqalign{B_{ll'}= \mathop{{\sum}'}_{i,j} {\,} \varphi^*_l({\bf R}_i) B({\bf R}_{ij}) M({\bf R}_i) M({\bf R}_j)
\varphi_{l'}({\bf R}_i), \\
B^{\dagger}_{ll'}= \mathop{{\sum}'}_{i,j} {\,} \varphi^*_l({\bf
R}_i) B^*({\bf R}_{ij}) M({\bf R}_i) M({\bf R}_j)
\varphi_{l'}({\bf R}_i)} \label{BHl}
\end{equation}

Matrices ${\cal L} = \left( \begin{array}{cc} A & -B^{\dagger}
\\ B & -A \end{array} \right)$ and ${\cal L'} = \left(
\begin{array}{cc} A & B \\ -B^{\dagger} &  -A \end{array}
\right)$ can be transformed into each other by replacing  $x$ with
$y$. Such a substitution formally replaces $B$ with $-B^{\dagger}$
and vise versa. Because the spectrum $E$ is invariant under such a
replacement, matrices ${\cal L}$ and ${\cal L'}$ should have the
same eigenvalues. From the other hand the similarity
transformation, $S^{-1} {\cal L} S$ with a matrix $S = \left(
\begin{array}{cc} 0 & -I \\ I & 0 \end{array} \right)$,
transforms ${\cal L}$ into $-{\cal L'}$. Thus, the eigenvalues of
${\cal L}$ should come up with pairs $E$ and $-E$.

The number of equations to solve can be substantially reduced if
we use the $w$- periodicity and
choose as a full set of function $\varphi_{\bf k}({\bf R}_i)=
1/\sqrt{N} \exp(i{\bf k R}_i)$. Then the system of $2N$ coupled
equations (\ref{ABm}), with $N$ being the number of sites of the
original cubic lattice, is reduced to $2N'$ equation, with
$N'= (w/a)^3$ being the number of sites within one $w$ period. The corresponding
Brillouin zone associated with $w$- periodicity has its boundaries
at $\pm \pi/w$ and reciprocal vectors ${\bf G}= 2\pi/w (l_x,
l_y, l_z)$, where $l_x$, $l_y$ and $l_z$ are integers. At a fixed
${\bf k}$ the operators (\ref{Al}) and (\ref{BHl}) mix
states ${\bf k}$ and ${\bf k+ G}$ which can differ at most by one
reciprocal vector. Number of different reciprocal
vectors in the domain $|{\bf G}| \leqslant 2\pi/a$ is $N'=
(w/a)^3$. The $2N \times 2N$ matrix ${\cal L}$ is now divided into
sub-blocks of a lower size $2N' \times 2N'$
\begin{equation} \label{Lpt}
{\cal L}=
\left(
\begin{array}{cccc}
{\cal L}({\bf k}_1) & & & \\
 & {\cal L}({\bf k}_2) & & \\
& & \ddots &
\end{array}
\right)
\end{equation}
Each sub-block ${\cal L}({\bf k})=
\left( \begin{array}{cc} A_{pt}({\bf k}) & -B^{\dagger}_{pt}({\bf k}) \\
B_{pt}({\bf k}) & -A_{pt}({\bf k}) \end{array} \right)$ is
labelled by a wave number ${\bf k}$, which runs over the first
Brillouin zone. The corresponding sub-block equation has the form
\begin{equation} \label{Apteq}
\left(
\begin{array}{cc}
A_{pt} & -B^{\dagger}_{pt} \\
B_{pt} & -A_{pt}
\end{array}
\right)
\left(
\begin{array}{c}
u_t \\
v_t
\end{array}
\right)
= E
\left(
\begin{array}{c}
u_p \\
v_p
\end{array}
\right)
\end{equation}
where we use the short-hand notation for matrix elements $A_{pt}=
A_{{\bf G}_p {\bf G}_t} ({\bf k})$ and eigenvectors
$\left(u_t \atop v_t \right)= (u_{G_1}({\bf k}), \dots, u_{G_{N'}}({\bf k}),
v_{G_1}({\bf k}), \dots, v_{G_{N'}}({\bf k}))^T$. In Eq.
(\ref{Apteq}) the sum is carried out over all
reciprocal lattice vectors ${\bf G}_t$. The coefficients $A_{pt}$ and $B_{pt}({\bf k})$
containing contributions from the anisotropy, exchange and dipolar
terms of the Hamiltonian (\ref{Htot}) are given by
\begin{eqnarray} \label{Apt}
\fl
A_{pt}({\bf k})= 2K_{\bf G_p- G_t} + \sum \limits_{\bf G} \widetilde J_{\bf G} \widetilde J_{\bf G_p- G_t- G} \left( F({\bf G})- F({\bf k+ G_p- G})\right) \nonumber \\
+ g\sum \limits_{\bf G} \bigl( A_1({\bf k- G})+ A_2({\bf G_t+ G}) \bigr)
M_{\bf G_p+G} M^*_{\bf G_t+G}
\end{eqnarray}
\begin{equation} \label{Bpt}
\eqalign{B_{pt}({\bf k})= g\sum \limits_{\bf G} B({\bf k-G})
M_{\bf G_p+G} M^*_{\bf G_t+G}, \\
B^{\dagger}_{pt}({\bf k})= g\sum \limits_{\bf G} B^*({\bf k-G})
M_{\bf G_p+G} M^*_{\bf G_t+G}.}
\end{equation}
Here $M_{\bf G}$ and $K_{\bf G}$ are Fourier coefficients of the
magnetization, $M({\bf R}_i)= \sum \limits_{\bf G}
M_{\bf G} e^{i{\bf GR}_i}$, and anisotropy, $K({\bf R}_i)=
\sum \limits_{\bf G} K_{\bf G} e^{i{\bf GR}_i}$, respectively. The
Fourier transform of exchange interaction $J_{ij}$ is given in
terms of $F({\bf k})= 2(\cos(k_x a)+ \cos(k_y a)+ \cos(k_z a))$
and the Fourier transform $\widetilde J_{\bf k}$ of an auxiliary
function $\widetilde J({\bf R}_i)$ which is equal to $J_h^{1/2}$
in the hard phase and $J_s^{1/2}$ in the soft phase. The exchange
coupling can be given in terms of $\widetilde J({\bf R}_i)$ as
$J_{ij}= \sum \limits_{\bf a} \widetilde J({\bf R}_i) \widetilde
J({\bf R}_j) \delta_{{\bf R}_i, {\bf R}_j+ {\bf a}}$, where the
summation is over nearest neighbors.

The lattice sums
$A_{1,2}({\bf k})= \mathop{{\sum}'}_{{\bf R}_{j}} {\:} e^{i{\bf k}
{\bf R}_{ij}} A_{1,2}({\bf R}_{ij})$ and $B({\bf k})=
\mathop{{\sum}'}_{{\bf R}_{j}} {\:} e^{i{\bf k} {\bf R}_{ij}}
B({\bf R}_{ij})$ can be presented in terms of the dipolar sum
$D^{\alpha \beta}({\bf k})$ as follows, $A_1({\bf k})= D^{zz}({\bf
k})/2$, $A_2({\bf G})= D^{zz}({\bf G})$ and $B({\bf k})= (-1/2)
\left( D^{xx}({\bf k})- D^{yy}({\bf k})- 2i D^{xy}({\bf k})
\right)$. The dipole sum $D^{\alpha \beta}({\bf k})$ is defined by
\begin{equation}
D^{\alpha \beta}({\bf k})= \mathop{{\sum}'}_{{\bf R}_{j}} {\:} e^{i{\bf k} {\bf R}_{ij}} D^{\alpha \beta}({\bf R}_{ij})
\end{equation}

To determine $D^{\alpha \beta}({\bf k})$ we use the Ewald
summation method \cite{Cohen55, Fujiki87}. As is known the tensor
$D^{\alpha \beta}({\bf k})$ is well- defined everywhere except
point ${\bf k}= 0$ \cite{Cohen55, Fujiki87}. Further we treat the
point ${\bf k} \to 0$ as the corresponding limit of $D^{\alpha
\beta}({\bf k})$. This limit depends on the direction of vector
${\bf k}$ and gives rise to the dependence of the spin wave
spectrum $E_{\bf k}$ on the direction of ${\bf k}$. The $D^{\alpha
\beta}(0)$ occurs only in one term, $A_2({\bf G})$. The value of
$D^{\alpha \beta}(0)$ is shape dependent and connected with the
demagnetizing field of the homogeneously magnetized
ellipsoidal-shaped magnetic body \cite{Akhiezer68}, $D^{\alpha
\beta}(0)= -4\pi N_{{\alpha}{\beta}}+ 4\pi/3 {\,}\delta_{\alpha
\beta}$. The magnetic charges forming on the outer surface of the
homogeneously magnetized finite body induce the demagnetizing field $H^{\alpha}_{dip}= -
N_{\alpha \beta} M^{\beta}_S$.

Here we are interested in
intrinsic properties of ferromagnetic media not affected by the
presence of magnetic charges forming on the outer boundary. For
infinitely extended system the demagnetizing tensor is
$N_{{\alpha}{\beta}}= 0$ \cite{Holstein40, Clogston56} and
$D^{\alpha \beta}(0)= (4\pi/3) {\,} \delta_{\alpha \beta}$. Earlier, to be consistent with the procedure of Monte Carlo simulation for a system with periodic boundary conditions we used $D^{\alpha \beta}(0)= 0$ \cite{ChuiJAP13}. In a homogeneous ferromagnetic media it would result in a downward shift of the spin wave spectrum relative to the spectrum of a ferromagnet free of the dipolar interaction.

The structure of the system of equations (\ref{Apteq}) is similar to the coupled Bogoliubov equations for bosonic excitations in the theory of a Bose condensate \cite{Fetter98}. The solution of the system of Eq. (\ref{Apteq}) yields in general a set of eigenfrequencies $E_j({\bf k})$ and eigenvectors $u_{tj}({\bf k}), v_{tj}({\bf k})$ labelled by a zone index $j$.
For real eigenvalues $E_j$ the corresponding eigenfunctions are normalized by the condition \cite{Fetter98}
\begin{equation} \label{norm}
\sum \limits_{t} \left( u_{tj'}^*({\bf k}) u_{tj}({\bf k})- v_{tj'}^*({\bf k}) v_{tj}({\bf k}) \right)= \delta_{jj'}.
\end{equation}
The spatial profile of these eigenvectors for a magnon with momentum ${\bf k}$
and zone index $j$ can be visualized with the help of magnon wave functions
$u_{{\bf k}j}({\bf R}_i)$ and $v_{{\bf k}j}({\bf R}_i)$, which can be found as follows:
\begin{equation}
\eqalign{u_{{\bf k}j}({\bf R}_i)= \frac{1}{\sqrt{N'}} \sum \limits_{{\bf G}_t}
e^{i{\bf G}_t{\bf R}_i} u_{tj}({\bf k}),
\\
v_{{\bf k}j}({\bf R}_i)=  \frac{1}{\sqrt{N'}} \sum \limits_{{\bf G}_t}
e^{i{\bf G}_t{\bf R}_i} v_{tj}({\bf k}).}
\end{equation}
They are normalized by the condition which is followed from Eq.
(\ref{norm})
\begin{equation} \label{normR}
\sum \limits_{{\bf R}} \left( |u_{{\bf k}j}({\bf R})|^2 - |v_{{\bf k}j}({\bf R})|^2 \right)= 1,
\end{equation}
where the summation is over one $w$- period.

\subsection{One-phase magnet}

The magnetostatic interaction modifies the spectrum of exchange
spin waves in two subtle ways. The first one is the dependence of
$E_{\bf k}$ on the direction of ${\bf k}$, which gives rise to the
notion of the spin wave manifold \cite{Stancil09}. The second one
is the dependence of $E_{\bf k}$ on the shape of the magnetic
body, it comes from the long-range nature of magnetostatic
forces. Before considering the effects of the magnetostatic
interaction on the spin wave spectrum of a composite, we briefly
illustrate the behavior of eigenfrequencies $E_j({\bf k})$ in a
one-phase magnet for comparison. It will serve as a basis for our
discussion of the spectrum for a two-phase magnet.

For the homogeneous one-component ferromagnet we can formally put
$w= a$ and there is only one reciprocal vector ${\bf G_p}= {\bf
G_t}= 0$ needs to be considered. In this case the system of
equations (\ref{Apteq}) reduces to two equations for amplitudes
$u_{\bf k}$ and $v_{\bf k}$ of the corresponding $u-v$
transformation. Accordingly, the corresponding coefficients
$A_{pt}({\bf k})$ and $B_{pt}({\bf k})$ reduced to coefficients
$A_{\bf k}$ and $B_{\bf k}$ of the Hamiltonian (\ref{H2}) given
below.

Spin waves in homogeneous ferromagnet is a well-studied problem
\cite{Holstein40, Akhiezer68, Kittel87}. In the harmonic
approximation the spin system Hamiltonian reads
\begin{equation} \label{H2}
H=  \sum \limits_{\bf k} {\,} \left \{ (\varepsilon_{\bf k}+ gA_{\bf k}) a^{\dagger}_{\bf k}
a_{\bf k} + \frac12 g B_{\bf k} a_{\bf k} a_{\bf k} + \frac12 g B^*_{\bf k} a^{\dagger}_{\bf k}
a^{\dagger}_{\bf k} \right \},
\end{equation}
where $\varepsilon_{\bf k}= 2K+ 6J(1- \gamma_{\bf k})$ is the
spectrum of the ferromagnet without the dipole-dipole interaction
term, $\gamma_{\bf k}= (1/3)(\cos(k_x a)+ \cos(k_y a)+ \cos(k_z
a))$, $A_{\bf k}= D^{zz}({\bf k})/2+ D^{zz}(0)$, $D^{zz}(0)=
\mathop{{\sum}'}_{{\bf R}_{j}} {\:} D^{zz}({\bf R}_{j})$ and
$B_{\bf k}$ coincides with $B({\bf k})$ given above. For the
case of infinitely extended ferromagnetic body when there is no
demagnetizing field coming from the boundary magnetic charges the
value of $D^{zz}(0)= 4\pi/3$.

\begin{figure}
\includegraphics[width=0.8\columnwidth]{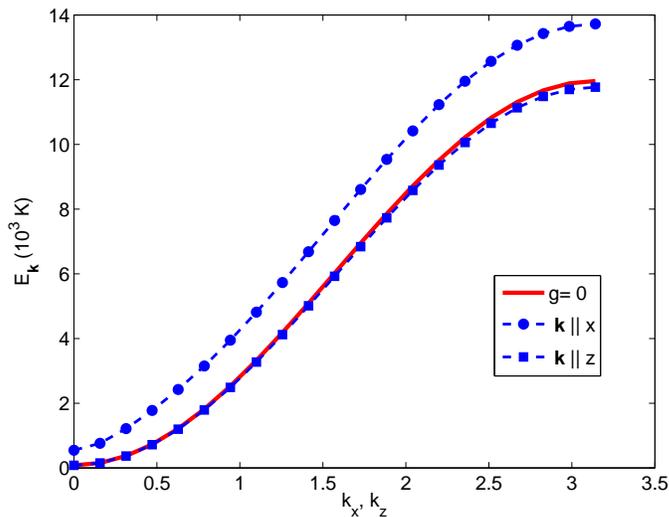}
\caption{\label{EFeCo} Spin wave excitations energies for the soft
phase FeCo. The spectrum (dashed lines) is split into the upper
and lower branches, corresponding to wave vectors parallel to the
$x$ and $z$ direction, respectively. The spectrum of the
ferromagnet $\varepsilon_{\bf k}$ without dipolar interaction ($g=
0$) is shown by solid line. }
\end{figure}

The eigenvalues of the Hamiltonian (\ref{H2}) is well known
\cite{Holstein40, Akhiezer68, Kittel87}
\begin{equation} \label{Ekh}
E_{\bf k}= \sqrt{(\varepsilon_{\bf k}+ gA_{\bf k})^2- g^2|B_{\bf k}|^2}
\end{equation}
They corresponds to spin waves with momentum ${\bf k}$ propagating
in infinitely extended magnetic body. For finite body this formula
holds true for $k \gg 1/L$, where $L$ is the size the magnetic
body. For this case  $D^{zz}(0)= -4\pi N_3+ 4\pi/3$ also accounts
for the demagnetizing field $H_{dip}= -N_3 M_S$. The regions $k
\lesssim 1/L$, corresponds to magnetostatic spin wave limit
\cite{Walker57}. The spectrum of these modes depends on the
boundary conditions of the magnetic body. We do not consider the
magnetostatic modes, and treat the point ${\bf k} \to 0$ as the
corresponding limit for the spin wave spectrum in an infinitely
extended magnetic sample.

We illustrate the dispersion relation (\ref{Ekh}) in Fig.
\ref{EFeCo} for the soft phase FeCo. We considered the case of
infinite media and thus $N_3= 0$ and $H_{dip}= 0$. Resulting spin wave excitations
lie within the spin wave manifold $E_{\bf k}$. The upper and the
lower branches  of the manifold correspond to the wave vectors
directed along $x$ and $z$ axes, respectively. The solid line
presents the spectrum $\varepsilon_{\bf k}$, which does not
account for the dipolar interaction. In the long wave limit, $ka \ll
1$, the lower branch of the manifold $E_{\bf k}$ coincides with
the spectrum of pure exchange spin waves $\varepsilon_{\bf k}= 2K+
D {\bf k}^2$, $D= Ja^2$ is exchange stiffness.

\section{Results and discussion}

We illustrate our numerical calculation for a hard- soft composite
SmCo$_5$/FeCo with the following parameters \cite{Coey10}:
$K_s/k_B= 40$ K, $J_s/k_B= 2.9 \cdot 10^3$ K, $\mu_0 M_s= 2.4$ T
and $g_{s}/k_B= \mu_0 M_s^2 v/(4 \pi k_B)= 290$ K for the soft phase
FeCo; $K_h/k_B= 13.3 \cdot 10^3$ K, $J_h/k_B= 3.3 \cdot 10^3$ K,
$\mu_0 M_h= 1.08$ T and $g_{h}/k_B= \mu_0 M_h^2 v/(4 \pi k_B)= 57$ K
for the hard phase SmCo$_5$, $k_B$ is the Boltzmann constant. The
effective lattice constant is $a= 2.2$ nm and is smaller than the
magnetic length of the hard phase $l_{m,h}= \pi(A_h a^3/K)^{1/2}
\simeq 2.5$ nm, $A_h$ is the exchange stiffness of the hard phase.
Calculation were performed for hard-soft periodicity $w/a= 20$.
Further it is convenient to rescale $E_j({\bf k})$ by a factor of
$k^{-1}_B$, so it is given in units of K. The linear dimensions of
the hard (soft) phase $l_h$ ($l_s$) will be given in units of the
cell size $a$ and the wave vector $k$ is rescaled via $ka$.

\begin{figure}
\includegraphics[width=0.75\columnwidth]{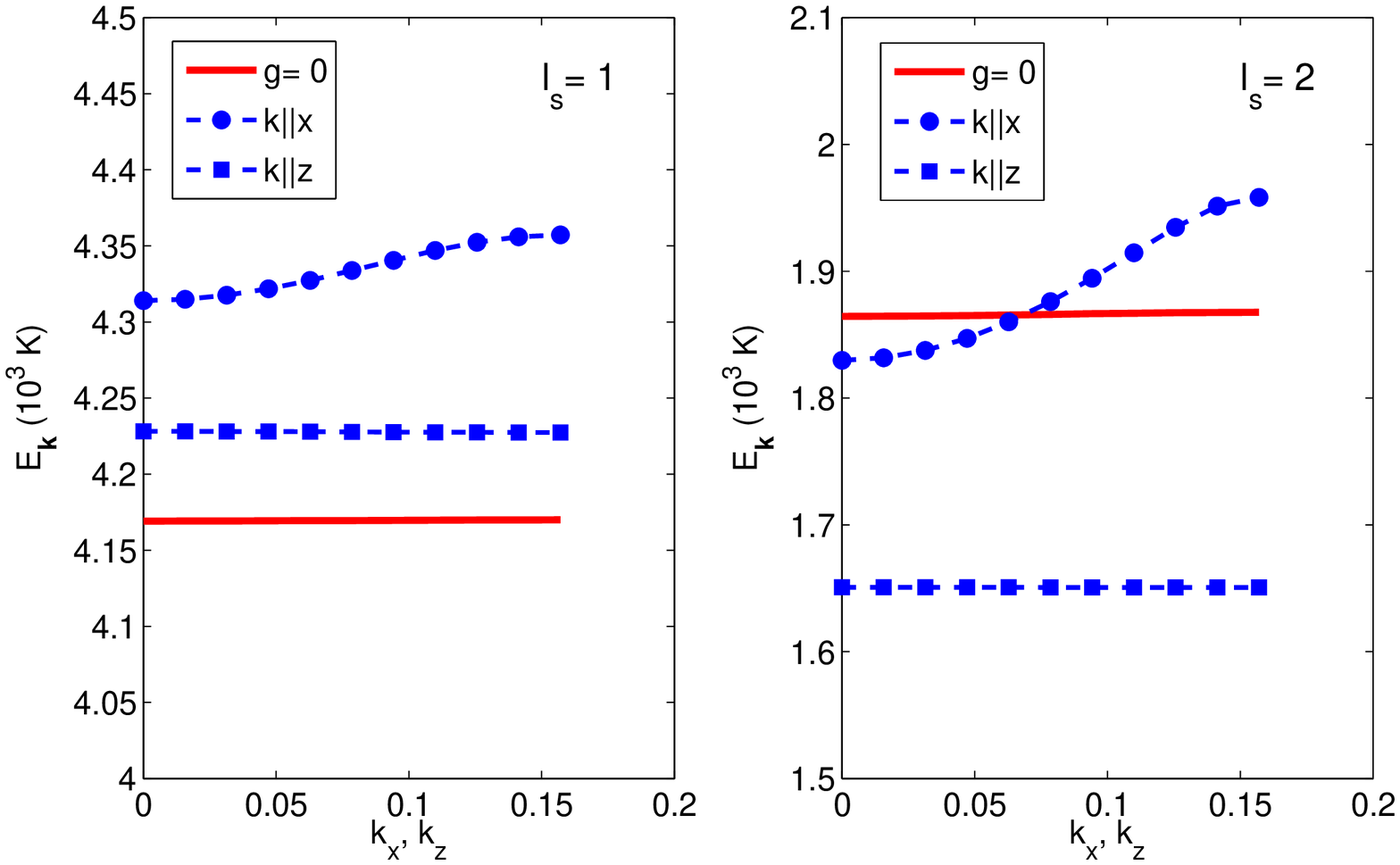}
\includegraphics[width=0.75\columnwidth]{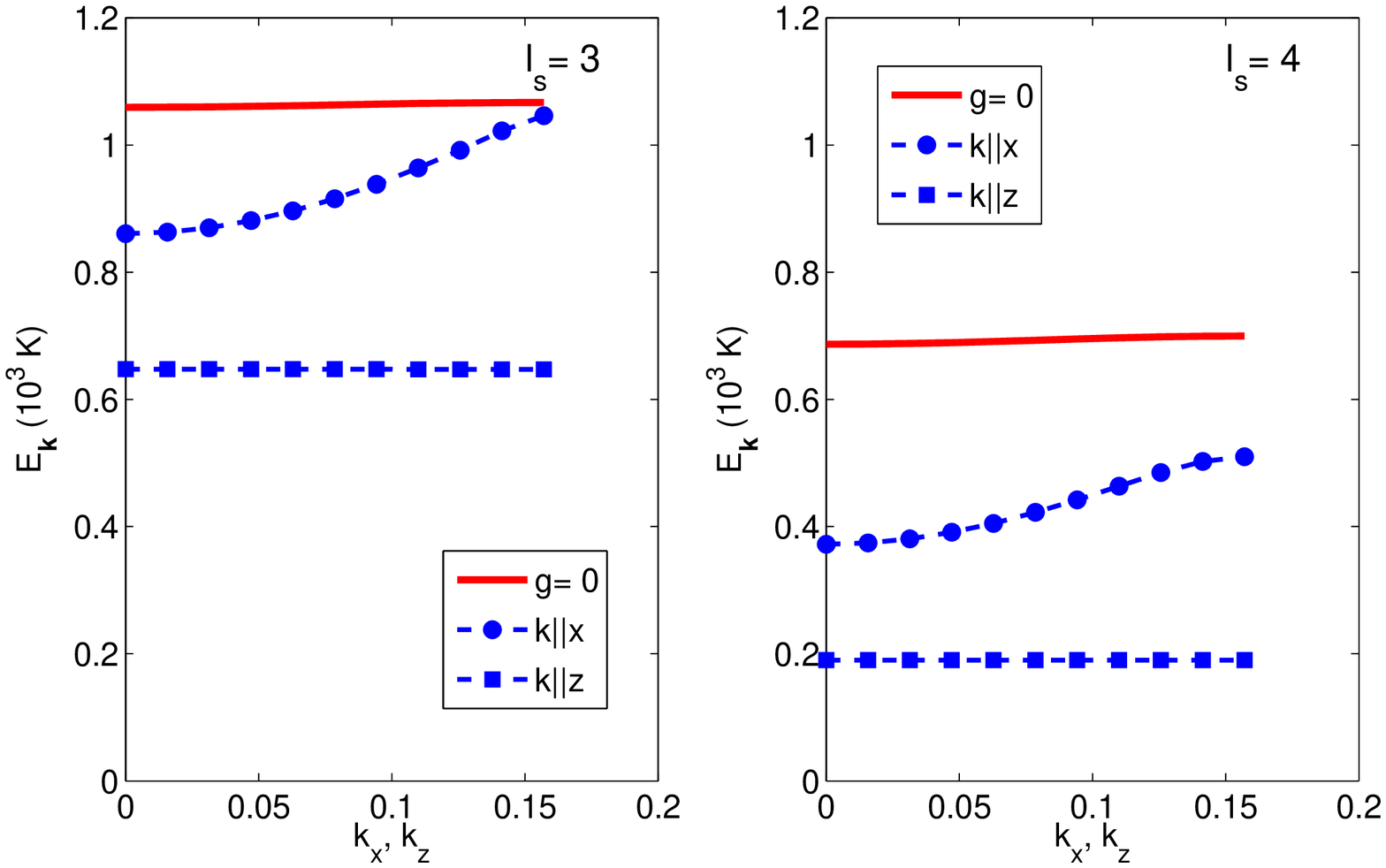}
\caption{\label{Els1} The spin wave spectrum evolution with
increasing soft phase content from $l_s= 1$ to $l_s= 4$. The solid
line ($g= 0$) represents result for the spectrum $\varepsilon_{\bf k}$ without the
dipole- dipole interaction. Lines with
markers represent the lower and upper branches of the spin wave
spectrum. }
\end{figure}

An example of the spin wave spectrum for the lowest magnon zone
resulting from numerical solution of eigenproblem (\ref{Apteq}) is
shown in Fig. \ref{Els1}. Four panels of Fig. \ref{Els1} show the
evolution of the spectrum with increasing soft phase content
$l_s$. The solid line presents the spin wave spectrum
$\varepsilon_{\bf k}$ obtained without the dipole-dipole
interaction term ($g= 0$). As expected the spin wave energies
depend on direction of the wave vector ${\bf k}$ forming the
spin-wave manifold. The lowest and the upper branches of the
spectrum correspond to the spin waves propagating parallel to the
$z$ and $x$ direction, respectively. The splitting is of order of
the characteristic magnetostatic energy of the soft phase $g_{s}
\simeq 300 K$. With adding more soft phase the spin wave manifold
gradually displaces downward relative to $\varepsilon_{\bf k}$.
This can be attributed to the existence of the nonzero effective
demagnetizing field due to the magnetic non-homogeneity. This is qualitatively different from
the spectrum shown at Fig. \ref{EFeCo}, with the lower branch of $E_{\bf k}$ matching $\varepsilon_{\bf k}$.
Eventually with increasing the soft phase amount the spin wave manifold presented
in Fig. \ref{Els1} touches $E_{\bf k}= 0$ signalling that the homogeneously magnetized state is no longer the ground state of the  periodic array of hard phase cubes immersed into the soft phase matrix. In this respect $E_{\bf k}= 0$ cannot gradually evolves into the manifold presented in Fig. \ref{EFeCo} with increasing the soft phase amount.

The corresponding magnon wave functions $u_{\bf k}({\bf R})$ and
$v_{\bf k}({\bf R})$, both for the upper and the lower
branches of $E_{\bf k}$, show the maximum magnitude in the soft phase and are strongly suppressed in the hard phase,
i.e. the low-lying spin excitations are mostly concentrated into the soft phase.
In our earlier work \cite{ChuiJAP13} we considered the hard-soft composite SmFeN/FeCo
with homogeneous exchange interaction $J_{ij}= J$. We found the the similar behavior of magnon wave functions
and the excitation spectrum $E_{\bf k}$ with increasing the soft phase content.
It indicates that the discontinuity of exchange constant $J_{ij}$ on hard-soft boundaries is not crucial for the aforementioned lowering of $E_{\bf k}$ with increasing $l_s$.
The situation closely parallels that studied
in Ref. \cite{Krawczyk08} for three-dimensional periodic structure
of two different soft ferromagnetic materials. They found that
the magnonic gap in spin wave dispersion is controlled by the
variation of the spontaneous magnetization contrast between two
phases.

The fluctuation of the magnetization is determined by the density of
the magnons which in turn is governed by the Boltzmann factor
$\exp(-E_j({\bf k})/k_BT)$. The spin wave energy of the composite
is controlled by the value of $K_s$ and the dipolar interaction.
The  amplitude of the lower (higher) lying excitations are
concentrated in the soft (hard) phase. The lower $K_s$ is in
comparison with $K_h$ the larger is the amplitude of the low-lying
excitation spin wave in the soft phase. For the experimental
values of small $K_s$ the spin excitations concentrate mostly in
the soft phase. The thermal excitation of these low lying magnons
gives rise to considerable fluctuations of the soft phase spins.
This leads to the lack of the remanence enhancement with
increasing of soft phase fraction, at it would expected from $M_r=
v_s M_s + v_h M_h$, where $v_s$ and $v_h$ are volume fractions of the
soft and hard phases. Even before the spin wave energy approaches
zero, this lowering will increase the finite temperature
fluctuation of the magnetization and lower the remanence $M_r$
that is approximately given by \cite{Kittel87}
\begin{equation}
M_r \simeq M_r(T= 0)- \frac{2}{N} \sum \limits_{\bf k} \frac{k_B
T}{E_{\bf k}}.
\end{equation}

\begin{figure}
\includegraphics[width=1.0\columnwidth]{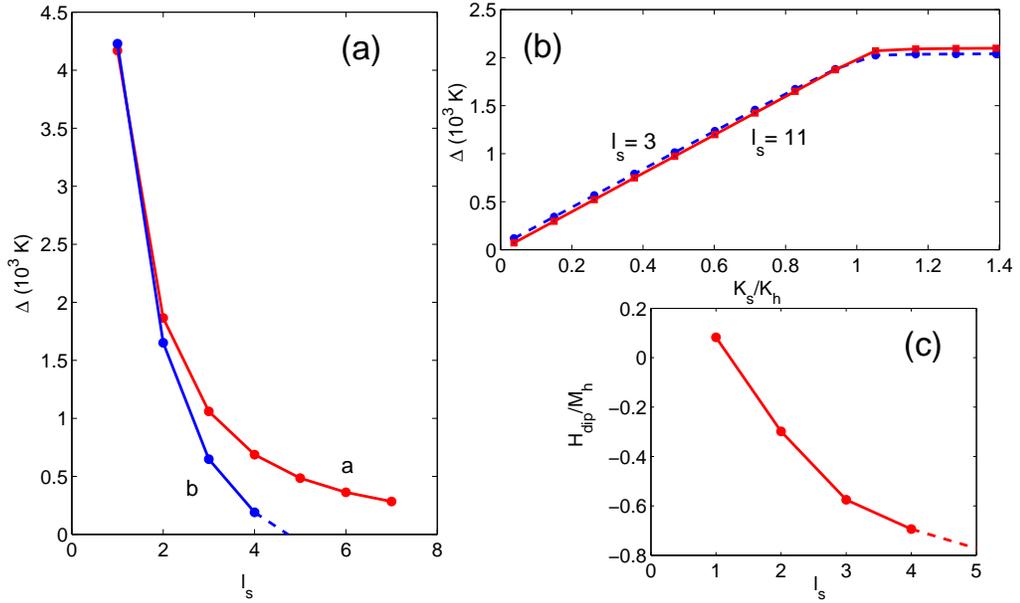}
\caption{ \label{Fig5a_plot}
(a) Spin wave gap $\Delta$ (panel ) as a function of the
the soft phase content $l_s$. The gap is calculated without (curve
$a$) and with (curve $b$) the dipolar interaction term.
(b) Spin wave gap $\Delta$ vs soft phase anisotropy parameter $K_s$ (normalized by $K_h$) for small ($l_s= 3$) and large ($l_s= 11$) amount of soft phase.
(c) The effective demagnetizing field $H_{dip}$ (normalized by $M_h$) vs soft phase content $l_s$. }
\end{figure}

To track the position of the lower branch of the spectrum we
introduce the magnonic gap $\Delta= \lim_{{\bf k} \to 0}  E_{\bf
k}$ in the long wave limit ${\bf k} \to 0$, for ${\bf k}$ parallel
to the $z$ direction. The effect of increasing soft phase content
on the spin wave gap is depicted in Fig. \ref{Fig5a_plot}(a) for system
without dipolar interaction (curve $a$) and with dipolar
interaction (curve $b$). Curves $a$ and $b$ present, respectively,
$\varepsilon_{\bf k}$ and the lower branch of $E_{\bf k}$ at ${\bf
k}= 0$. With adding more soft phase curve $a$ tends to the finite
value of $\Delta= 2K_s$, corresponding to the energy of long wave
excitations of the soft phase, while curve $b$ tends to zero. At
some critical value $l_{s,c} \simeq 5$ the gap closes, and the
system in fully aligned state becomes unstable. It is signalled by the appearance
of complex eigenvalues in the spectrum of the matrix, Eq. (\ref{Apteq}).

The energy difference between curves $a$ and $b$ can be attributed
to the internal demagnetizing field arising from nonhomogeneous
behavior of $M_S({\bf r})$ and the corresponding magnetic charges
formed at hard-soft boundaries. In the absence of the
demagnetizing effect curves $a$ and $b$ would coincide. To
introduce the effective demagnetizing field let us invoke to the
spectrum of one-phase ferromagnet in the long wave limit $1/L \ll
k \ll 1/a$. In this limit the tensor $D^{\alpha \beta}({\bf k})$
has the form \cite{Akhiezer68} $D^{\alpha \beta}({\bf k})= -4\pi
k_{\alpha}k_{\beta}/k^2+ 4\pi/3 {\,} \delta_{\alpha \beta}$ and
the coefficients $A_{\bf k}= 2\pi \sin^2 \theta_{\bf k}$ and
$|B_{\bf k}|= 2\pi \sin^2 \theta_{\bf k}$, where $\theta_{\bf k}$
is the angle between ${\bf k}$ and $z$ axis. The spin wave
dispersion has the well known form \cite{Holstein40, Akhiezer68, Kittel87}
\begin{equation}
E_{\bf k}= \sqrt{(\varepsilon_{\bf k} - 4\pi N_3 g)( \varepsilon_{\bf k}- 4\pi N_3 g+ 4\pi g \sin^2 \theta_{\bf k})}
\end{equation}
The lower branch of the manifold $E_{\bf k}$ corresponds to spin
waves travelling parallel to the $z$ direction, $\theta_{\bf k}=
0$. If one is interested in properties of an infinite magnet,
when there is no demagnetizing field \cite{Holstein40,
Clogston56}, then  $N_3= 0$, and respectively, $H_{dip}= 0$. In
this case the lower branch of $E_{\bf k}$ coincides with the
spectra of pure exchange spin waves $\varepsilon_{\bf k}= 2K+ D
{\bf k}^2$, as it was presented in Fig. \ref{EFeCo} above. On the other hand, if there are magnetic charges (as, for example, for a finite homogeneously magnetized ferromagnetic body) then the lower branch,
$E_{\bf k}= \varepsilon_{\bf k} - 4\pi N_3 g$, is shifted downward
by the amount proportional to the internal dipolar field $H_{dip}=
-N_3 M_S$.

In the case of nonhomogeneous magnet the magnetic charges forming on internal hard-soft boundaries
produce the intrinsic dipolar field $H_{dip}$, which affects the spin-wave spectrum of the magnet.
It shifts the whole manifold $E_{\bf k}$ downward respective to the $\varepsilon_{\bf k}$.
One can estimate the internal dipolar field from the dependence $E_{\bf k}$ as
\begin{equation}
E_{\bf k}= \varepsilon_{\bf k}+ \frac{H_{dip}}{M_h}4\pi g
\end{equation}

The effective demagnetizing field $H_{dip}$ is presented in Fig. \ref{Fig5a_plot}(c). As expected, with increasing soft phase
content $l_s$ the magnitude of the internal field increases due to
increase of a magnetic charge forming at the hard-soft boundary.
As a consequence, the whole manifold $E_{\bf k}$ displaces
downward respective to $\varepsilon_{\bf k}$. When the dipolar
field increases there is some critical amount of the soft phase
$l_{s,c}$, beyond which a system with magnetic or structural
nonhomogeneity would undergo the phase transition into another
ground state with non-homogeneous distribution of the
magnetization.

We finally present the dependence of $\Delta$ on the value of soft phase anisotropy parameter, $K_s$.
The presented range of $K_s$ is much larger than that is in real soft magnets and was chosen to demonstrate the character of scaling of $\Delta$ with $K_s$.
This dependence is illustrated in panel (b) of Fig. \ref{Fig5a_plot} for small, $l_s= 3$, and large, $l_s= 11$, amount of soft phase. The spin gap demonstrates a linear dependence on $K_s$, as long as $K_s < K_h$. It suggest that the main mechanism in formation of low-energy spin excitations in composite magnets is the interplay of the anisotropy and exchange interaction, as in the case of pure exchange spin waves, $\varepsilon_{\bf k}= 2K+ D{\bf k}^2$. The role of the dipolar interaction amounts to shifting of the spin-wave manifold as a whole and is crucial in the vicinity of $K_s \to 0$. This is the case for SmCo$_5$/FeCo composite, where $K_s \ll K_h$ and, as we have shown above, the spin excitations are mostly concentrated in the soft phase.

\section{Conclusion}

In this paper we studied the spin-wave spectra of two-phase
ferromagnetic composites. Hard phase cubes inserted into a soft phase matrix form a 3D
structure of exchange-coupled hard and soft phases.
In such a geometry there are regions where the saturation
magnetization $M_S({\bf r}) {\hat z}$ is perpendicular to the
hard-soft boundary, which is energetically unfavorable from the
magnetostatic point of view. We have found the corresponding spin
wave excitations and magnon wave functions by diagonalizing the
Hamiltonian in the harmonic approximation.

We numerically found that low-lying excitations are mostly
concentrated in the soft phase, which gives rise to considerable
fluctuation of soft phase spins and a lack of the remanence
enhancement with increasing of the soft phase amount. The spin wave
frequencies strongly depend on the magnetic charges and
magnetostatic interaction in the system. As it turns out the difference in the exchange coupling
constants of hard and soft phase has little effect, if any, on the eigenmodes of the system. The energy
of the lowest spin wave zone is mostly affected by the presence of the effective demagnetizing field
in the composite and the value of anisotropy energy of the soft phase.

Finally, we have shown that with adding more soft phase the fully
aligned spin state ${\bf M}({\bf r})= M_S({\bf r}) {\hat z}$
becomes unstable relative to excitation of long wavelength spin
waves with wave vector ${\bf k}$ parallel to the $z$ direction.
The system undergoes a phase transition into another ground state
characterized by misalignment of spins of hard and soft phases. We
have found critical values of soft phase thickness $l_{s,c}$.

We thank G. Hadjipanayis and A. Gabay for helpful discussions. This work was supported in part by the U.S. Department of Energy
(DOE) under the ARPA-E program (S.T.C.).

\section*{References}

\end{document}